\documentclass[11pt, onecolumn]{article}

\usepackage{fancyhdr}
\usepackage[english]{babel}
\usepackage{abstract}

\usepackage{amsmath}  
\usepackage{amssymb} 
\usepackage{mathrsfs}

\usepackage{euscript}
\usepackage{hyperref}

\usepackage{cyr}
\usepackage{epsfig}
\usepackage{epstopdf}

\usepackage{titlesec}

\makeatletter
\renewcommand{\@oddhead}{}
\renewcommand{\@oddfoot}{\hfill ---~\thepage~---\hfill}
\makeatother

\titleformat*{\section}{\Large\bfseries}
\titleformat*{\subsection}{\large\bfseries}

\fancypagestyle{firststyle} 
{
\fancyhead[L]{\small Published in Astronomy Letters, 2020, Vol. 46, No.7, pp. 449-461 \\ original russian text: Pis'ma v Astronomicheskii Zhurnal, 2020, Vol. 46, No. 7, pp. 480-493.\hfill}
\fancyfoot[L]{\hfill ---~\thepage~---\hfill}
 
}

\begin{document}
\thispagestyle{firststyle}

\begin{center}
\Large Collisional Pumping of OH Masers near Supernova Remnants

\vspace{0.5cm}
\large A.V. Nesterenok
\vspace{0.5cm}

\normalsize Ioffe Physical-Technical Institute, Politekhnicheskaya St. 26, Saint~Petersburg, 194021 Russia

e-mail: alex-n10@yandex.ru
\end{center}

\begin{abstract} 
\noindent
The collisional pumping of OH masers in non-dissociative C-type shocks near supernova remnants is considered. The emergence of maser emission in OH lines is investigated for various shock parameters -- the shock speed, the preshock gas density, the cosmic-ray ionization rate, and the magnetic field strength. The largest optical depth in the 1720 MHz line is reached at high gas ionization rates $\zeta \geq 10^{-15}$~s$^{-1}$, an initial density $n_{\rm H,0} \leq 2 \times 10^4$~cm$^{-3}$, and a shock speed $u_{\rm s} \geq 20$~km~s$^{-1}$. According to our calculations, there is also a level population inversion for the 6049 and 4765~MHz transitions of excited OH rotational states. However, the optical depth in these lines is small for all of the investigated shock parameters, which explains the non-detection of maser emission in these lines in supernova remnants.
\end{abstract}

Keywords: \textit{cosmic masers, shocks, supernovae.}
\smallskip

DOI: 10.1134/S1063773720070075

\section*{Introduction}
OH maser emission is observed for all transitions of the ground rotational state of the molecule $^2\Pi_{3/2}$ $j = 3/2$ -- the main 1665 and 1667~MHz lines and the satellite 1612 and 1720~MHz lines. As a rule, the maser emission in the 1665 and 1667~MHz lines is associated with star-forming regions, while the masers in the 1612~MHz line are associated with late-type stars (Caswell 1998; Qiao et al. 2020). Masers in the 1720~MHz line are observed predominantly near supernova remnants (SNRs) and in star-forming regions (Beuther et al. 2019). The 1720~MHz masers near SNRs have the following differences from the masers in this line near HII regions in star-forming regions: the maser clumps have a larger size and a lower luminosity, the emission has a relatively low degree of circular and linear polarization ($\lesssim 10\%$), and the magnetic fields in the emission sources are weaker by several times (Caswell 1999; Hoffman et al. 2005b; Brogan et al. 2013). Furthermore, other lines of the ground OH rotational state are observed in absorption (Hewitt et al. 2006). About 10\% of the SNRs in our Galaxy have 1720 MHz OH maser emission sources (Brogan et al. 2013). The 1720~MHz maser emission is considered as a signature of the interaction of a SNR with a molecular cloud (Frail et al. 1994; Frail and Mitchell 1998; Wardle and Yusef-Zadeh 2002).

Elitzur (1976) was the first to show that the 1720 ~MHz maser has the collisional pumping mechanism. The sizes of the maser regions near supernovae and the theoretical estimates of the physical conditions needed for efficient OH maser pumping suggest that the 1720~MHz maser emission is generated in non-dissociative magnetohydrodynamic C-type shocks (Lockett et al. 1999) (this type of shocks is described below). High OH column densities, $N_{\rm OH} \simeq 10^{16}-10^{17}$~cm$^{-2}$, are needed for the generation of maser emission (Lockett et al. 1999; Wardle and McDonnell 2012). Based on a simple shock model, Wardle (1999) showed that the gas ionization rate should be $\sim 10^{-15}$~s$^{-1}$ for the formation of the required number of OH molecules downstream of the shock. Both enhanced cosmic-ray fluxes and X-ray emission from the SNR-filling hot gas can be responsible for the gas ionization in molecular clouds near SNRs (Yusef-Zadeh et al. 2003; Hewitt et al. 2009; Schuppan et al. 2014; Phan et al. 2020). Energetic electrons are produced during the interaction of cosmic-ray particles and/or X-ray emission with molecular gas. The collisional excitation of H$_2$ molecules by energetic electrons and the subsequent H$_2$ emission in the Lyman and Werner bands is the source of ultraviolet (UV) radiation in the molecular gas (Prasad and Tarafdar 1983). OH molecules are produced in H$_2$O photodissociation reactions and ion-molecule reactions.

As a rule, the cloud model in which the physical parameters -- the gas density, the column density of emitting molecules, and the temperature -- are independent and vary in a wide range is used to model the maser pumping (see, e.g., Lockett et al. 1999). The intensity of radiation in OH maser lines depends mainly on the postshock column density of molecules, which depends on the shock extent and the number density of OH molecules. The shock length (in the direction of propagation) is inversely proportional to the gas density and ionization rate, while high gas ionization rates are needed for the formation of OH molecules downstream of the shock. Thus, the physical parameters -- the shock length and the number density of OH molecules -- are not independent. The physical shock model imposes constraints on the range of physical parameters, which should be taken into account when modelling the maser pumping. In our calculations we use the shock model published in Nesterenok (2018) and Nesterenok et al. (2019). The goal of this paper is to determine the shock parameters needed for efficient collisional OH maser pumping.

\section*{The C-type shock model}
If the relative speed of the colliding gas flows is higher than any propagation speed of perturbations in the interstellar medium, then a J-type (from the word 'jump') shock is formed. In this case, the physical parameters at the shock front change abruptly and the kinetic energy of the gas flows dissipates in a small region. This leads to strong gas heating and the dissociation of molecules. If the speed of the gas flows is lower than the magnetosonic speed, but higher than the speed of sound for the neutral gas component, then there is a compression of the ionic gas component and the magnetic field upstream of the shock (Mullan 1971; Draine 1980). The shock front is smeared due to the scattering of ions by neutral particles and the parameters of the neutral gas component undergo smooth changes. As a result, a C-type (from the word 'continuous') shock is formed. The H$_2$ molecule is the main coolant of the molecular gas. Therefore, the C-type shocks can exist for speeds less than some limiting value at which a complete dissociation of H$_2$ molecules occurs at the shock front. For gas densities and ionization fractions typical of dark molecular clouds, the limiting speeds of C-type shocks are approximately $40-60$~km~s$^{-1}$ (Le Bourlot et al. 2002; Nesterenok et al. 2019).

The model of a steady-state C-type shock propagating in a dense molecular cloud was proposed in Nesterenok (2018) and Nesterenok et al. (2019). Below we list the main physical processes that are taken into account in the model:

(1) A steady flow of partially ionized gas is considered, with the magnetic field direction being perpendicular to the gas velocity direction. The magnetic field lines are assumed to be frozen into the ionic gas component. A small mismatch between the velocities of the ionic and neutral gas components is specified at the coordinate origin, which increases along the gas flow. The velocities, densities, and temperatures of the gas components (ions with electrons and neutral particles) as functions of coordinates are determined by integrating the mass, momentum, and energy conservation equations (Roberge and Draine 1990). The numerical integration of the equations stops as soon as the velocity difference between the ions and the neutral gas component drops below some specified low value.

(2) The model takes into account a complete chemical reaction network -- the gas-phase chemical reactions, the adsorption and desorption of chemical species on dust grains, and the chemical reactions on the surface of dust grains. The gas-phase chemical reaction network was taken from the UDfA 2012 database (McElroy et al. 2013). The photodissociation and photoionization reaction rates were updated according to the data from Heays et al. (2017). The collisional molecule dissociation reactions are also taken into account (Nesterenok 2018). The chemical reaction network on the surface of dust grains was taken from the NAUTILUS code (Ruaud et al. 2016). The model takes into account the sputtering of icy grain mantles in the hot gas at the shock front (Draine and Salpeter 1979).

(3) The system of differential equations includes the equations for the energy level populations of CI, CII, OI ions and H$_2$, CO, H$_2$O molecules, which is needed to calculate the gas cooling rates. The Sobolev approximation (the approximation of a high velocity gradient) is used to calculate the intensity of radiation in spectral lines (Sobolev 1957; Hummer and Rybicki 1985).

(4) The model takes into account the main gas cooling and heating processes -- the gas heating due to the flow of the neutral gas component and ions through each other and the gas cooling due to the radiation in atomic and molecular lines. The gas heating due to the exothermic chemical reactions and the photoelectric effect on dust as well as the heating by cosmic-ray particles are also taken into account.

(5) The dust is assumed to be composed of spherical silicate particles 0.05~$\mu m$ in radius. The ratio of the masses contained in the dust and gas per unit volume of the interstellar gas is taken to be 0.01. The model takes into account the main processes of electric charge acquisition and neutralization by dust grains -- the photoelectric effect, the attachment of electrons, and the neutralization of ions on dust grains. The velocity of charged dust grains is close to the velocity of the ionic gas component (Draine 1980). For gas densities $n_{\rm H_2} \gtrsim 10^4-10^5$~cm$^{-3}$ the momentum is transferred between the neutral gas component and ions mainly through the scattering of neutral atoms and molecules by charged dust grains. The length of the shock front is inversely proportional to the initial gas density and ionization rate.

The shock modelling consists of two parts: (1) modelling the chemical evolution of a dark molecular cloud and (2) modelling the shock propagation. The age of the molecular cloud is assumed to be $5 \times 10^5$~yr. In our calculations the gas ionization rate is assumed to be constant during the chemical evolution of the cloud. Note that in molecular clouds near SNRs this parameter can depend on time, which can affect the chemical composition of the gas (Nesterenok 2019). The empirical relation $B_0 = \beta n^{1/2}_{\rm H,0}$, where $n_{\rm H,0}$ is the total number density of hydrogen nuclei, approximately holds for the magnetic fields in molecular clouds (Crutcher 1999). Observations of the Zeeman effect for the 1720 MHz line allowed the magnetic field in OH maser sources to be estimated, $B \simeq 0.5-2$~mG (Brogan et al. 2000, 2013; Hoffman et al. 2005a, 2005b). The measured magnetic field strengths in OH masers roughly correspond to the mean value in the interstellar medium (Brogan et al. 2000). Table 1 gives the shock parameters that were used in our calculations.

~\\
~\\
\begin{tabular}{l@{\quad\quad}|@{\quad\quad}l}
\multicolumn{2}{l}{\large\bf Table 1. Shock parameters} \\ [5pt]
\hline \\ [-2ex]
Preshock number density of hydrogen nuclei, $n_{\rm H,0}$ & $2 \times 10^3 - 2 \times 10^5$~cm$^{-3}$ \\ [5pt]
Shock speed, $u_{\rm s}$ & $5-60$~km~s$^{-1}$ \\ [5pt]
Cosmic-ray ionization rate, $\zeta$ & $10^{-16} - 3 \times 10^{-15}$~s$^{-1}$ \\ [5pt]
Initial ortho-H$_2$ to para-H$_2$ ratio & 0.1 \\ [5pt]
Magnetic field strength, $\beta$ & 1 \\ [5pt]
Visual extinction, $A_{\rm V}$ & 10 \\ [5pt]
Turbulent velocity, $v_{\rm turb}$ & 0.3~km~s$^{-1}$ \\ [5pt] \hline
\end{tabular}
~\\
~\\
For the total number density of hydrogen nuclei we have $n_{\rm H,0} = n_{\rm H} + 2 n_{\rm H_2}$, where $n_{\rm H}$ and $n_{\rm H_2}$ are the number densities of hydrogen atoms and molecules, respectively; the parameter $\beta$ characterizes the magnetic field strength, $B [\rm \mu G] = \beta \left( n_{\rm H,0} [\text{\rm cm}^{-3}] \right)^{1/2}$.
~\\

\section*{Calculation of the OH energy level populations}
\subsection*{The collisional rate coefficients and spectroscopic data}
The spin--orbit coupling of the unpaired electron in the O atom leads to the separation of the OH rotational levels into two subsets, $^2\Pi_{1/2}$ and $^2\Pi_{3/2}$. Each OH rotational level experiences $\Lambda$-doubling, with each $\Lambda$-doublet sublevel having an opposite total parity. The energy levels with parities $(-1)^{j-1/2}$ and $-(-1)^{j-1/2}$ are denoted by $e$ and $f$, respectively (Marinakis et al. 2019). Each of the $\Lambda$-doublet sublevels experiences hyperfine splitting. The hyperfine splitting levels differ by the quantum number of the total angular momentum $F$. The OH level structure leads to a difference in the frequencies of some transitions between various rotational states of the molecule by a few MHz, which is comparable to the thermal line broadening (Burdyuzha and Varshalovich 1973). In solving the radiative transfer equation in OH lines, it is important to take into account the line overlap in frequency (Litvak 1969; Doel et al. 1990).

In our calculations we took into account 56 OH energy levels (with hyperfine splitting). The energy of the highest level $^2\Pi_{1/2}$, $j = 6.5e$, $F = 7$ is 1550 K. The spectroscopic data for OH were taken from the HITRAN 2016 database (Gordon et al. 2017). The collisional rate coefficients for OH--H$_2$ collisions were calculated by Offer et al. (1994) for 24 lower OH energy levels. The collisional rate coefficients for collisions between OH molecules and He atoms were obtained by Marinakis et al. (2016, 2019) for 56 lower OH energy levels. The maximum gas temperature $T_{\rm max}$ for which the rate coefficients were calculated is 200 and 300 K for the OH--H$_2$ and OH--He data, respectively. For gas temperatures exceeding $T_{\rm max}$ we used the collisional rate coefficients at $T_{\rm max}$. The data for OH--H$_2$ collisions are accessible in the LAMDA database (Sch\"{o}ier et al. 2005); the data for OH--He collisions were kindly provided by Dr. Marinakis.

\subsection*{The system of statistical equilibrium equations for the energy level populations}

The shock profile obtained through our numerical simulations is broken down into layers and the OH energy level populations are computed for each layer. In our calculations statistical equilibrium is assumed to hold for the OH energy level populations $(dn/dt = 0)$. This is true when the time scales of the change in physical parameters in the shock are much greater than the relaxation time scales of the level populations. This condition may not be fulfilled at the shock peak, where the physical parameters of the gas change rapidly. However, for the cooling postshock gas the assumption about statistical equilibrium of the energy level populations is valid (Flower and Gusdorf 2009).

The system of equations for the level populations is

\begin{equation}
\begin{array}{c}
\displaystyle
\sum_{k=1, \, k \ne i}^M \left( R_{ki}(z)+C_{ki} \right) n_k(z) - n_i(z)\sum_{k=1, \, k \ne i}^M \left( R_{ik}(z)+C_{ik} \right)=0, \quad i=1,...,M-1, \\
\displaystyle
\sum_{i=1}^M n_i(z)=1,
\end{array}
\label{eq5}
\end{equation}

\noindent
where $M$ is the total number of energy levels, $R_{ik}(z)$ is the probability of the transition from level i to level k due to the radiative processes, and $C_{ik}$ is the probability of the transition from level $i$ to level $k$ due to the collisional processes. The radiative transition probabilities $R_{ik}(z)$ are

\begin{equation}
\begin{array}{c}
\displaystyle
R_{ik}^{\downarrow}(z)=B_{ik}J_{ik}(z)+A_{ik}, \quad \varepsilon_i > \varepsilon_k, \\[10pt]
\displaystyle
R_{ik}^{\uparrow}(z)=B_{ik}J_{ik}(z), \quad \varepsilon_i < \varepsilon_k,
\end{array}
\end{equation}

\noindent
where $A_{ik}$ and $B_{ik}$ are the Einstein coefficients for spontaneous and stimulated emission, $J_{ik}(z)$ is the intensity of radiation averaged over the direction and line profile, and $\varepsilon_i$ is the energy of level $i$. The emission of inverted transitions is neglected in our calculations of the energy level populations.

The system of equations for the energy level populations (1) is solved by the method of iterations. In each step we calculate the mean intensities of radiation in spectral lines based on the level populations derived in the previous step. Next, using the system of equations (1), we find new values of the level populations. The convergence criterion for the derived series of energy level populations is the condition on the population increment in one step:

\begin{equation}
\displaystyle 
\max_i |\Delta n_i/n_i| < 10^{-5}.
\end{equation}

\noindent
This error in the energy level populations corresponds to the error in the gain $\Delta \gamma < 10^{-18}$~cm$^{-1}$ for the 1720~MHz OH line (for a number density of OH molecules $n_{\rm OH} = 1$~cm$^{-3}$ and Doppler line broadening $v_{\rm D} > 0.1$~km~s$^{-1}$).

\subsection*{The Sobolev approximation with dust absorption and overlapping spectral lines}

In this section we present a generalization of the method for calculating the mean intensities in spectral lines described in Hummer and Rybicki (1985) to the case of an overlap of two spectral lines.

In a one-dimensional geometry the intensity of radiation $I$ at frequency $\nu$ depends on the depth $z$ and the angle $\theta$ between the $z$ axis and the direction of the radiation. The quantity $\mu = {\rm cos} \theta$ and the dimensionless frequency $x$ are commonly used instead of the variable $\theta$ and the frequency $\nu$, respectively:

\begin{equation}
\displaystyle
x = \frac{\nu - \nu_{ik}}{\Delta \nu_{\rm D}},
\end{equation}

\noindent
where $\nu_{ik}$ is the mean line frequency, $\Delta \nu_{\rm D}$ is the line profile width, $i$ and $k$ are the numbers of the upper and lower energy levels of an atom or a molecule. The line width is determined by the thermal velocity of emitting molecules (atoms) $v_{\rm T}$ and the turbulent velocity in the gas--dust cloud $v_{\rm turb}$:

\begin{equation}
\displaystyle
\Delta\nu_{\rm D}=\nu_{ik}\frac{v_{\rm D}}{c}, \quad v_{\rm D}^{2}=v_{\rm T}^2+v_{\rm turb}^2,
\end{equation}

\noindent
where $v_{\rm T} = \sqrt{2kT_{\rm g}/m}$, $T_{\rm g}$ is the kinetic gas temperature. 

Consider the problem of radiative transfer in a spectral line without any overlap in frequency with other lines. The radiative transfer equation in a medium with a specified velocity field $v(z)$ can be written as

\begin{equation}
\displaystyle
\mu \frac{dI(z,\mu,x)}{dz} = \left[-\kappa_{\rm L}(z) I(z,\mu,x) + \varepsilon_{\rm L}(z) \right] \phi\left[x - \mu\frac{v(z)}{v_{\rm D}}\right] - \kappa_{\rm C} I(z,\mu,x) + \varepsilon_{\rm C},
\label{eq_radtransf1}
\end{equation}

\noindent
where $\varepsilon_{\rm C}$ and $\kappa_{\rm C}$ are the continuum emission and absorption coefficients, respectively, $\varepsilon_{\rm L}(z)$ and $\kappa_{\rm L}(z)$ are the line emission and absorption coefficients, respectively, averaged over the line profile; $\phi(x) = (\pi)^{-1/2}\text{exp}(-x^2)$ is the normalized spectral line profile (with this notation the absorption coefficient at the line center is $\kappa_{\rm L}/\sqrt{\pi}$). The additional term in the argument of the line profile in Eq. (\ref{eq_radtransf1}) takes into account the Doppler frequency shift when passing to a moving reference frame.

In the Sobolev approximation the length scale of the change in physical parameters in the medium is assumed to be much greater than the size of the resonance region $\Delta z_{\rm S}$, where the radiation at a given frequency interacts with the medium,

\begin{equation}
\displaystyle
\Delta z_{\rm S} = v_{\rm D} \left\vert \frac{dv}{dz} \right\vert^{-1}.
\label{eq_sobolev_condition}
\end{equation}

\noindent
Let us introduce the following parameters:

\begin{equation}
\displaystyle
\gamma = \frac{1}{\kappa_L v_{D}}\frac{dv}{dz}, \quad \delta = \frac{1}{\kappa_C v_{D}}\frac{dv}{dz}.
\end{equation}

\noindent
The reciprocals of these parameters, $1/\vert \gamma \vert$ and $1/\vert\delta \vert$, are equal to the line and continuum optical depths at a distance $\Delta z_{\rm S}$, respectively.

Lockett et al. (1999) showed that the effect of dust radiation on the collisional pumping of the OH 1720~MHz maser becomes significant at a dust temperature $T_{\rm d} \gtrsim 50$~K. In our calculations the dust temperature $T_{\rm d}$ does not exceed 40~K at the shock peak and is about 10~K in the region where the OH maser emission emerges. Below we will neglect the dust radiation. In this case, the intensity of radiation in a spectral line averaged over the frequency and direction can be calculated from the formula (Hummer and Rybicki 1985):

\begin{equation}
\displaystyle
J(z) = S_{\rm L}(z) \left[1 - 2\mathcal{P}(\delta, \gamma) \right], 
\label{eq_line_intens1}
\end{equation}

\noindent
where $S_{\rm L} = \varepsilon_{\rm L}/\kappa_{\rm L}$ is the source function in the spectral line, $\mathcal{P}(\delta, \gamma)$ is the probability of photon escape through one of the cloud boundaries or absorption by dust. For $\kappa_{\rm C} \to 0$, $\delta \to \infty$ the function $\mathcal{P}(\delta, \gamma)$ transforms to a well-known expression (Sobolev 1957; Hummer and Rybicki 1985):

\begin{equation}
\displaystyle
\mathcal{P}(\infty, \gamma) = \frac{\vert \gamma \vert}{2} \int\limits_0^1 d\mu \, \mu^2 \left[1 - {\rm exp} \left(-\frac{1}{\vert \gamma \vert \mu^2}\right)\right]
\end{equation}

\noindent
The values of $\mathcal{P}(\delta, \gamma)$ were calculated in Nesterenok (2016) for a wide range of parameters $\gamma$ and $\delta$. 

Consider the problem of radiative transfer in the case of two closely spaced (in frequency) spectral lines. The radiative transfer equation is

\begin{equation}
\begin{array}{c}
\displaystyle
\mu \frac{dI(z,\mu,x)}{dz} = \left[-\kappa_{\rm L1}(z) I(z,\mu,x) + \varepsilon_{\rm L1}(z) \right] \phi\left[x - \mu\frac{v(z)}{v_{\rm D}}\right] \\ [10pt]
\displaystyle
+ \left[-\kappa_{\rm L2}(z) I(z,\mu,x) + \varepsilon_{\rm L2}(z) \right] \phi\left[x + \Delta x - \mu\frac{v(z)}{v_{\rm D}}\right] - \kappa_{\rm C} I(z,\mu,x) + \varepsilon_{\rm C},
\end{array}
\label{eq_radtransf2}
\end{equation}

\noindent
where $\Delta x = (\nu_{1} - \nu_{2})/\Delta \nu_{\rm D}$ is the relative shift in the mean line frequencies. The lines are assumed to be closely spaced in frequency, $\vert\nu_{1} - \nu_{2}\vert \ll \nu_{1}$. If $\vert\Delta x\vert > 1$, then there is a nonlocal line overlap -- the radiation in the first line emitted in one region of space is superimposed in frequency on the radiation in the second line in the neighboring region located at a distance about $\vert\Delta x\vert \, \Delta z_{\rm S}$. By analogy with (\ref{eq_line_intens1}), for the intensity of radiation in the first line we can write

\begin{equation}
\displaystyle
J_{1}(z) = S_{\rm L1}(z) \left[1 - 2\mathcal{P}_{11}(\delta, \gamma_{1}, \gamma_{2}, \Delta x) \right] + S_{\rm L2}(z) \left[1 - 2\mathcal{P}_{12}(\delta, \gamma_{1}, \gamma_{2}, \Delta x) \right],
\label{eq_line_intens2}
\end{equation}

\noindent
where the first term is equal to the contribution of the radiative transitions between the first pair of energy levels to the radiation intensity, while the second term is equal to the contribution of the radiative transitions between the second pair of energy levels. For $\mathcal{P}_{11}$ and $\mathcal{P}_{12}$ the following expressions hold (Nesterenok 2020):

\begin{equation}
\begin{array}{c}
\displaystyle
\mathcal{P}_{11}(\delta, \gamma_{1}, \gamma_{2}, \Delta x) = \frac{1}{2} - \frac{1}{2} \int\limits_0^1 \frac{d\mu}{\mu^2\vert\gamma_1\vert} \int\limits_{-\infty}^{\infty} dx \, \phi(x)\int\limits_{x}^{\infty} dx' \phi(x') \times \\ [10pt] 
\displaystyle
\times \text{exp} \left\lbrace -\frac{1}{\mu^2\vert\gamma_1\vert} \int\limits_{x}^{x'} du \left[ \phi(u) + \frac{\gamma_1}{\gamma_2} \phi(u \mp \Delta x) + \frac{\gamma_1}{\delta} \right] \right\rbrace, \\ [10pt]
\displaystyle
\mathcal{P}_{12}(\delta, \gamma_{1}, \gamma_{2}, \Delta x) = \frac{1}{2} - \frac{1}{2} \int\limits_0^1 \frac{d\mu}{\mu^2\vert\gamma_2\vert} \int\limits_{-\infty}^{\infty} dx \, \phi(x \mp \Delta x) \int\limits_{x}^{\infty} dx' \phi(x') \times \\[10pt] 
\displaystyle
\times \text{exp} \left\lbrace -\frac{1}{\mu^2\vert\gamma_1\vert} \int\limits_{x}^{x'} du \left[ \phi(u) + \frac{\gamma_1}{\gamma_2} \phi(u \mp \Delta x) + \frac{\gamma_1}{\delta} \right] \right\rbrace,
\end{array}
\label{eq_loss_prob_func2}
\end{equation}

\noindent
where the upper and lower signs in the argument of the line profile correspond to positive and negative velocity gradients, respectively. The expression to calculate the intensity of radiation in the second line is derived from Eq. (\ref{eq_line_intens2}) as a result of the substitutions $S_{L1} \leftrightarrow S_{L2}$, $\gamma_1 \leftrightarrow \gamma_2$ and $\Delta x \leftrightarrow -\Delta x$. The integrals (13) were calculated in Nesterenok (2020) for a grid of parameters $\gamma_1$, $\gamma_2/\gamma_1$, $\Delta x$ and $\delta$. The ranges of values for these parameters were chosen to be the following: $\gamma_1$ -- $[10^{-6},10^{6}]$, $\gamma_2/\gamma_1$ -- $[10^{-2},10^{2}]$, $\Delta x$ -- $[-4,4]$, $\delta$ -- $[10^{2},10^{6}]$. The integration was done with the algorithms published in Press et al. (2007), the relative integration error was taken to be $10^{-5}$.

For each group of OH lines closely spaced in frequency (the transitions between the hyperfine splitting doublets of various rotational states) the overlap of the line profiles was considered only for pairs of lines close in frequency. The simultaneous overlap of three lines was not considered.

\section*{Results}
\subsection*{The shock structure and the number density of OH molecules}

\begin{figure}[h]
\centering
\includegraphics[width = 1.0\textwidth]{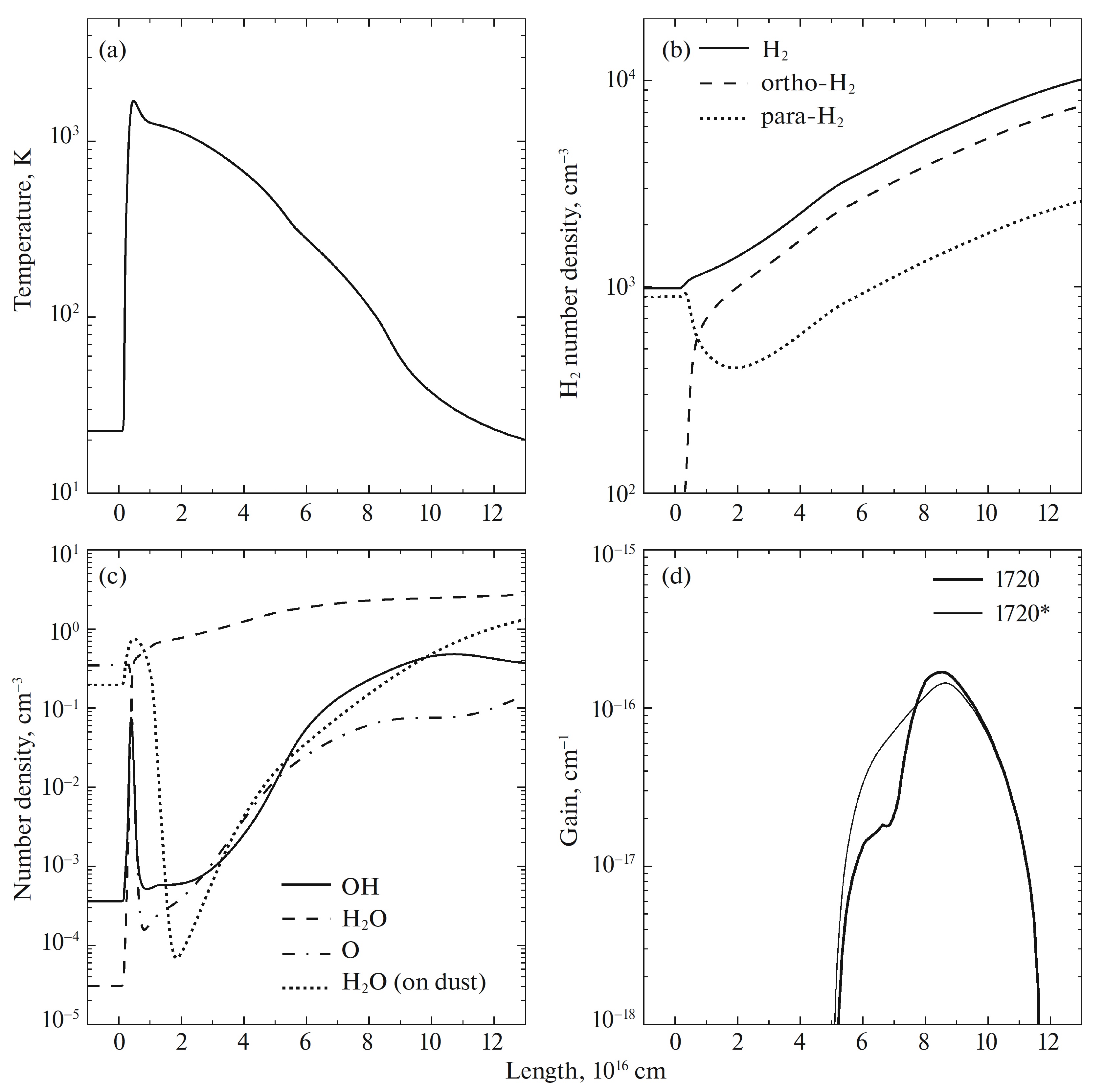}
\caption{Physical parameters in a C-type shock versus distance: (a) the temperature of the neutral gas component; (b) the number density of H$_2$ molecules; (c) the number density of O atoms, OH and H$_2$O molecules in the gas, and H$_2$O molecules on dust grains; (d) the gain at the line center for OH maser transitions. The shock parameters are the preshock gas density $n_{\rm H,0} = 2 \times 10^3$~cm$^{-3}$, the shock speed $u_{\rm s} = 20$~km~s$^{-1}$, and the cosmic-ray ionization rate $\zeta = 10^{-15}$~s$^{-1}$. Panel (d) also presents the results of our calculations in which the OH line overlap was neglected in the Sobolev approximation (marked by the asterisk).}
\end{figure}

\begin{figure}[h]
\centering
\includegraphics[width = 1.0\textwidth]{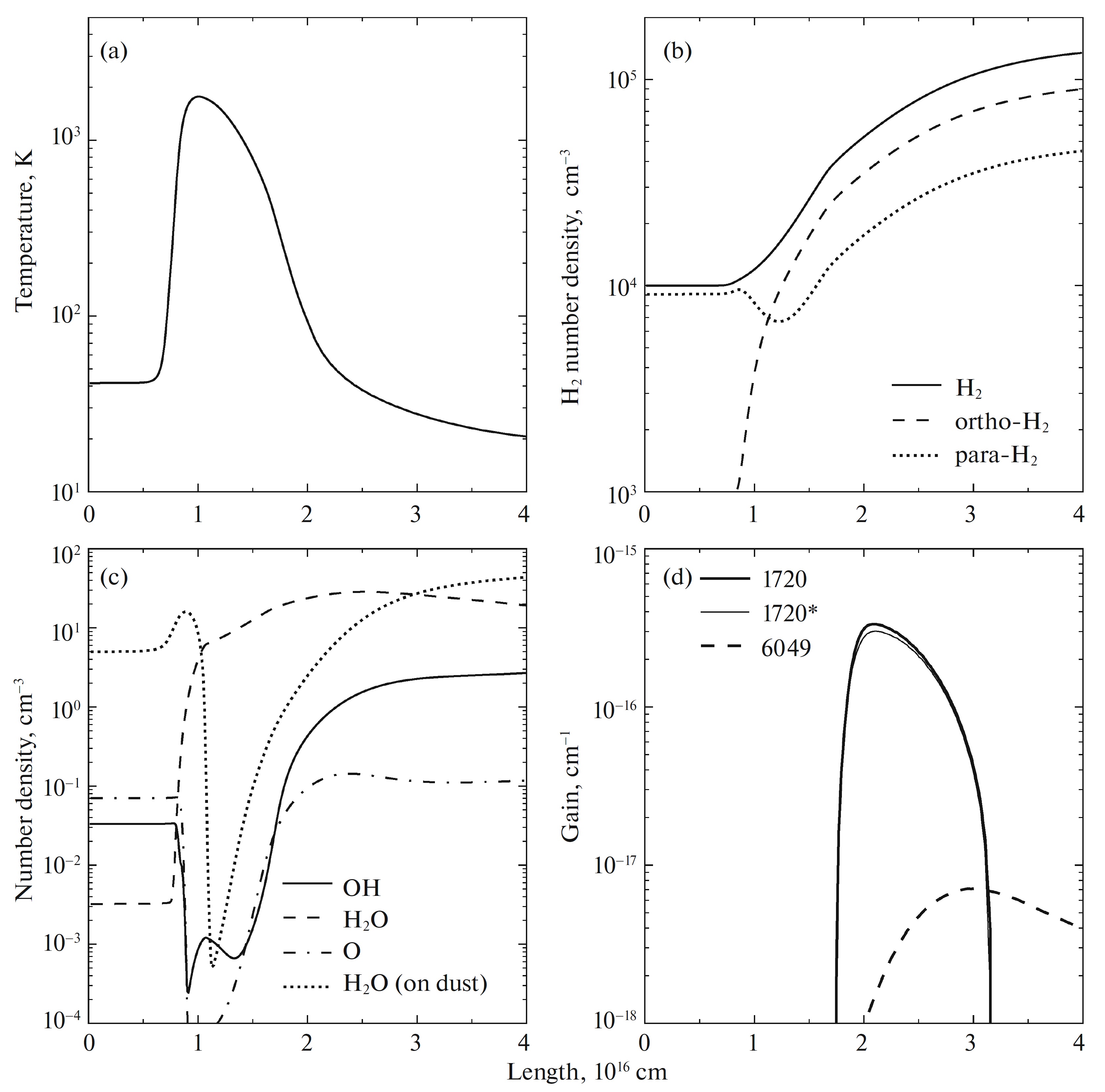}
\caption{Same as Fig. 1, but for the preshock gas density $n_{\rm H,0} = 2 \times 10^4$~cm$^{-3}$. The calculations in which the line overlap was taken into account in the Sobolev approximation and the calculations in which this effect was neglected yield close results for the 6049~MHz line.}
\end{figure}

\begin{figure}[h]
\centering
\includegraphics[width = 1.0\textwidth]{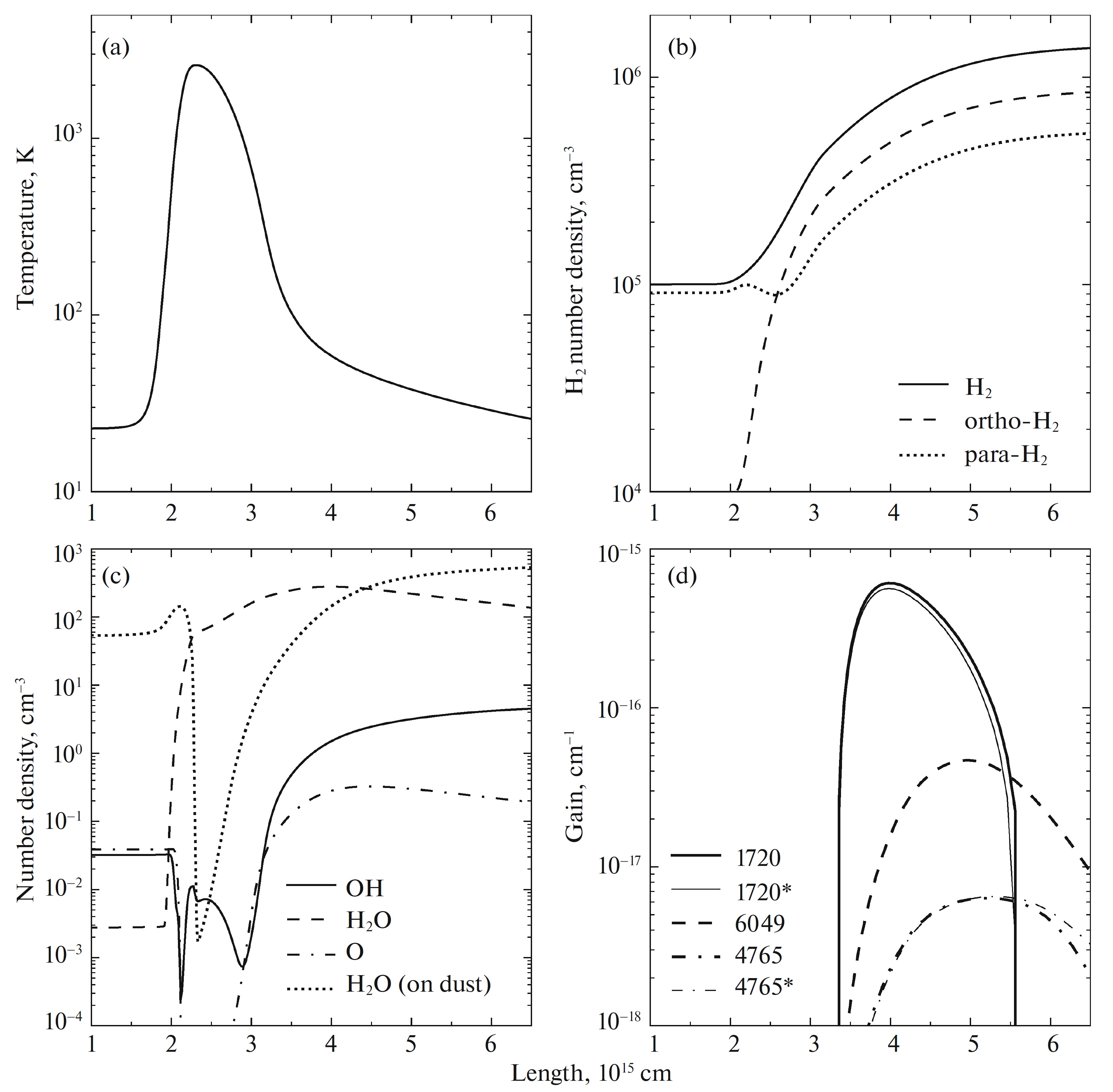}
\caption{Same as Fig.~1, but for the preshock gas density $n_{\rm H,0} = 2 \times 10^5$~cm$^{-3}$.}
\end{figure}

In Figs. 1--3 the temperature of the neutral gas component, the number densities of O atoms and H$_2$, OH, and H$_2$O molecules, and the gain in OH maser lines in a C-type shock are plotted against the distance. The results are presented for three preshock gas densities, $n_{\rm H,0} = 2 \times 10^3$, $2 \times 10^4$, and $2 \times 10^5$~cm$^{-3}$, while $\zeta = 10^{-15}$~s$^{-1}$ and $u_{\rm s} = 20$~km~s$^{-1}$. 

The initial ortho- to para-H$_2$ ratio in the preshock gas was assumed to be 0.1. The 'reactive' (spin changing) collisions of H$_2$ molecules with H atoms are the main mechanism for the conversion of para-H$_2$ to ortho-H$_2$ in the shock (Lique et al. 2014; Nesterenok et al. 2019). At gas ionization rates $\zeta = 10^{-15}$~s$^{-1}$ the preshock number density of H atoms is quite high (a few 10~cm$^{-3}$), which provides an efficient para- to ortho-H$_2$ conversion in the hot gas at the shock front. The ortho- to para-H$_2$ ratio in the postshock gas is about 3, 2, and 1.6 for initial densities $n_{\rm H,0} = 2 \times 10^3$, $2 \times 10^4$, and $2 \times 10^5$~cm$^{-3}$, respectively (Figs. 1b--3b). In this case, ortho-H$_2$ is the main collisional partner of the OH molecule. Pavlakis and Kylafis (1996, 2000) showed that the 1720 and 6049~MHz line population inversion is sensitive to the ortho- to para-H$_2$ ratio. This is due to the following property of the collisional rate coefficients calculated by Offer et al. (1994): the collisions of OH with para-H$_2$ reduce the upper energy level population for the 1720~MHz maser transition more efficiently than the lower level population (Lockett et al. 1999).

During the chemical evolution of a molecular cloud the atomic oxygen contained in the gas is consumed in chemical reactions. At low densities the chemical evolution of the cloud is slow and the gas-phase atomic oxygen can be the main reservoir of oxygen in the molecular gas (Fig. 1c). At high gas densities the H$_2$O molecules in the icy mantles on dust grains are the main reservoir of oxygen (Figs. 2c and 3c). At the shock peak, when the relative speed of the ions (along with the dust grains) and the neutral gas component is maximal, the sputtering of the icy mantles on the surface of dust grains occurs. At the same time, the gas-phase atomic oxygen turns into H$_2$O in chemical reactions. The fractional abundance of H$_2$O molecules in the gas increases by several orders of magnitude and reaches about $(2-3) \times 10^{-4}$ relative to the hydrogen nuclei (Figs. 1c--3c).

In the postshock region, after the evaporation of the icy mantles of dust grains, the main channel of H$_2$O production and OH destruction is the reaction

\begin{equation}
\displaystyle
{\rm H_2 + OH \to H_2O + H}.
\label{eq_h2o_prod}
\end{equation}
 
\noindent
This reaction proceeds rapidly at gas temperatures $T_{\rm g} \simeq 1000$~K or higher -- the activation energy is $E_{\rm a} \simeq 1700$~K (UDfA 2012; McElroy et al. 2013). The reverse reaction has an activation energy $E_{\rm a} \simeq 10^4$~K and its rate is much lower than the rate of reaction (\ref{eq_h2o_prod}). Therefore, the reaction equilibrium is shifted toward H$_2$O (Figs. 1c--3c).

The main OH production channels in the cooling gas are the photodissociation of H$_2$O molecules by UV radiation and the dissociative recombination reactions involving H$_3$O$^+$ ions:

\begin{equation}
\begin{array}{c}
\displaystyle
{\rm H_2O + \text{\rm UV CR} \to OH + H}, \\ [15pt]
\displaystyle
{\rm H_3O^+ + e^- \to OH + H + H}, \\ [15pt]
\displaystyle
{\rm H_3O^+ + e^- \to OH + H_2}.
\end{array}
\label{eq_oh_prod}
\end{equation}

\noindent
These reactions lead to an increase in the postshock number density of OH molecules as soon as the gas temperature drops below 1000~K (Figs. 1c--3c). The ion--molecule reactions initiated by cosmic-ray particles are the sources of H$_3$O$^+$ ions in dark molecular clouds. Thus, the higher the cosmic-ray ionization rate, the higher the postshock number density of OH molecules.

The OH production rate in reactions (15) is proportional to the H$_2$O number density, which, in turn, is proportional to the gas density. The OH adsorption rate on dust grains is also proportional to the gas density (the main OH destruction channel in the gas phase at high densities). Therefore, the number density of OH molecules in the shock tail hardly changes as the density $n_{\rm H,0}$ increases from $2 \times 10^4$ to $2 \times 10^5$~cm$^{-3}$ and is about $2-3$~cm$^{-3}$ in the case under consideration (Figs. 2c and 3c).

\subsection*{Applicability of the Sobolev approximation}
For the Sobolev approximation to be applicable, the length scale of the change in physical parameters $l$ must satisfy the condition $l \gg \Delta z_{\rm S}$. As an estimate of $l$ we may take (Gusdorf et al. 2008)

\begin{equation}
\displaystyle
l \sim T_{\rm g}/(dT_{\rm g}/dz)
\end{equation}

\noindent
In the cooling postshock gas $l \simeq (3 - 5) \Delta z_{\rm S}$. It makes no sense to take into account the line overlap for a line frequency difference $\vert\Delta x\vert > l/\Delta z_{\rm S}$, because for such $\Delta x$ the Sobolev approximation ceases to hold and the expressions for the photon loss probabilities (\ref{eq_loss_prob_func2}) become invalid.

In the region downstream of a C-type shock, where the OH maser emission is generated, the typical values of the velocity gradient for the neutral gas component are about $10^{-11}-10^{-10}$~cm~s$^{-1}$~cm$^{-1}$ for a preshock gas density of $2 \times 10^{4}$~cm$^{-3}$, ionization rate $\zeta = 10^{-15}$~s$^{-1}$, and a shock speed $u_{\rm s} = 20$~km~s$^{-1}$. The size of the region along the $z$ axis, where the OH maser emission emerges, is a few $\Delta z_{\rm S}$. The column density of OH molecules at distance $\Delta z_{\rm S}$ in the direction of shock propagation is $n_{\rm OH}\Delta z_{\rm S} \sim 10^{15}$~cm$^{-2}$ -- it is this parameter that determines the radiative transfer in molecular lines. The absolute values of the parameter $\gamma$ for OH transitions lie in a wide range, from about $10^{-2}$ (for the transitions between the states $^2\Pi_{3/2}$ $j = 5/2$ and $j = 3/2$) and higher. The typical values of the parameter $\delta$ are $10^3 - 10^4$ for infrared lines.

\subsection*{Maser emission in OH lines}
The expression for the gain at the center of a $i \to k$ line is

\begin{equation}
\displaystyle
\gamma_{ik}(z)=\frac{\lambda^2 A_{ik} n_{\rm OH}}{8 \pi \sqrt{\pi} \, \Delta \nu_{\rm D}} \left(n_i(z)-\frac{g_i}{g_k}n_k(z) \right),
\end{equation}

\noindent
where $g_i$ and $g_k$ are the statistical weights of the levels. Let us introduce a parameter $\tau_{\rm eff}$ -- the effective optical depth in the line in the direction of shock propagation,

\begin{equation}
\displaystyle
\tau_{\rm eff} = \int\limits_{0}^{\infty} \, dz \, \gamma_{ik}(z)
\end{equation}

\noindent
In the latter expression the frequency shift of the line profile is neglected and the integration is over the region with $\gamma_{ik} > 0$. The optical depth along the line of sight at an angle $\theta$ to the direction of shock propagation close to $90^{\circ}$ is approximately equal to $\tau_{\rm eff}/{\rm cos}\theta$ (the frequency shift of the line profile is negligible for such directions of the line of sight).

Figures 1d--3d show the gain at the line center for OH maser transitions as a function of the postshock distance. The energy level population inversion for the 1720~MHz line emerges when the gas temperature drops to $150-400$~K (depending on the initial gas density). The level population inversion vanishes as soon as the gas temperature drops below 30~K. It is these temperatures that were considered previously as the most favourable ones for the generation of OH maser emission in the 1720~MHz line (Elitzur 1976; Pavlakis and Kylafis 1996; Lockett et al. 1999). According to our calculations, there is an energy level population inversion for the transitions belonging to the following excited rotational states of the molecule: $^2\Pi_{3/2}$, $(j\varepsilon,F) = (2.5f,3) \to (2.5e, 2)$ at 6049~MHz and $^2\Pi_{1/2}$, $(j\varepsilon,F) = (0.5f,1) \to (0.5e, 0)$ at 4765~MHz. The 6049 and 4765~MHz lines are analogues of the satellite 1720~MHz line (the transitions between the extreme upper and lower sublevels of the rotational state). There is an energy level population inversion for the 1612 and 1665~MHz transitions of the ground rotational state at some shock parameters, but the optical depth is small.

The line overlap suppresses the population inversion for the 1720~MHz transition at high temperatures for $n_{\rm H,0} = 2 \times 10^3$~cm$^{-3}$ (Fig. 1d). In the remaining cases, the calculations in which the OH line overlap is taken into account and the calculations in which this effect is neglected give close values for the gain in the 1720~MHz maser line -- a difference is about $10\%$. This effect was considered previously by Wardle and McDonnell (2012). They showed that the energy level population inversion in the 1720 and 6049~MHz lines is suppressed if the local line overlap for a line profile broadening $v_{\rm D} > 0.5$~km~s$^{-1}$ is taken into account.

Figure 4 presents the results of our calculations of the effective optical depth in OH maser lines in the direction of shock propagation. The largest optical depth in the 1720~MHz line is reached at high gas ionization rates $\zeta \geq 10^{-15}$~s$^{-1}$ and an initial density $n_{\rm H,0} \leq 2 \times 10^4$~cm$^{-3}$. The results of our calculations confirmed the conclusions by Wardle (1999) that high gas ionization rates are needed for the generation of OH maser emission. The effective optical depth $\tau_{\rm eff}$ is about $3-5$. The higher the gas density, the smaller the length of the shock front. As a result, the optical depth in the 1720~MHz OH maser line is smaller for $n_{\rm H,0} = 2 \times 10^5$~cm$^{-3}$ than that in the case with a lower $n_{\rm H,0}$.

For $\zeta = 10^{-15}$~s$^{-1}$ we performed additional calculations in which the preshock magnetic field strength was assumed to be twice as high, $\beta = 2$. When the magnetic field doubles, the width of the shock front increases approximately by a factor of 2, while the maximum gas temperature decreases, other things being equal. The size of the postshock region, where the physical conditions are favourable for the pumping of OH masers, is larger in the case of a higher magnetic field. As a result, the optical depth in maser lines is higher by a factor of $2-3$. The sputtering of icy grain mantles and the release of H$_2$O occur at higher shock speeds. Therefore, the dependence of the optical depth on the shock speed is shifted rightward (Fig. 4).

For gas ionization rates $\zeta = 10^{-16}$~s$^{-1}$ the optical depth in the 1720~MHz line is $\tau_{\rm eff} < 1$. At this ionization rate the OH number density in the shock tail is $\sim 0.1$~cm$^{-3}$, which is responsible for the low gains. At gas ionization rates $\zeta \leq 10^{-16}$~s$^{-1}$ and shock speeds $u_{\rm s} \lesssim 20-25$~km~s$^{-1}$ the para- to ortho-H$_2$ conversion in the shock is quite inefficient. In this case, the main collisional partner of the OH molecule is para-H$_2$, which additionally reduces the energy level population inversion.

\begin{figure}[h]
\centering
\includegraphics[width = 0.91\textwidth]{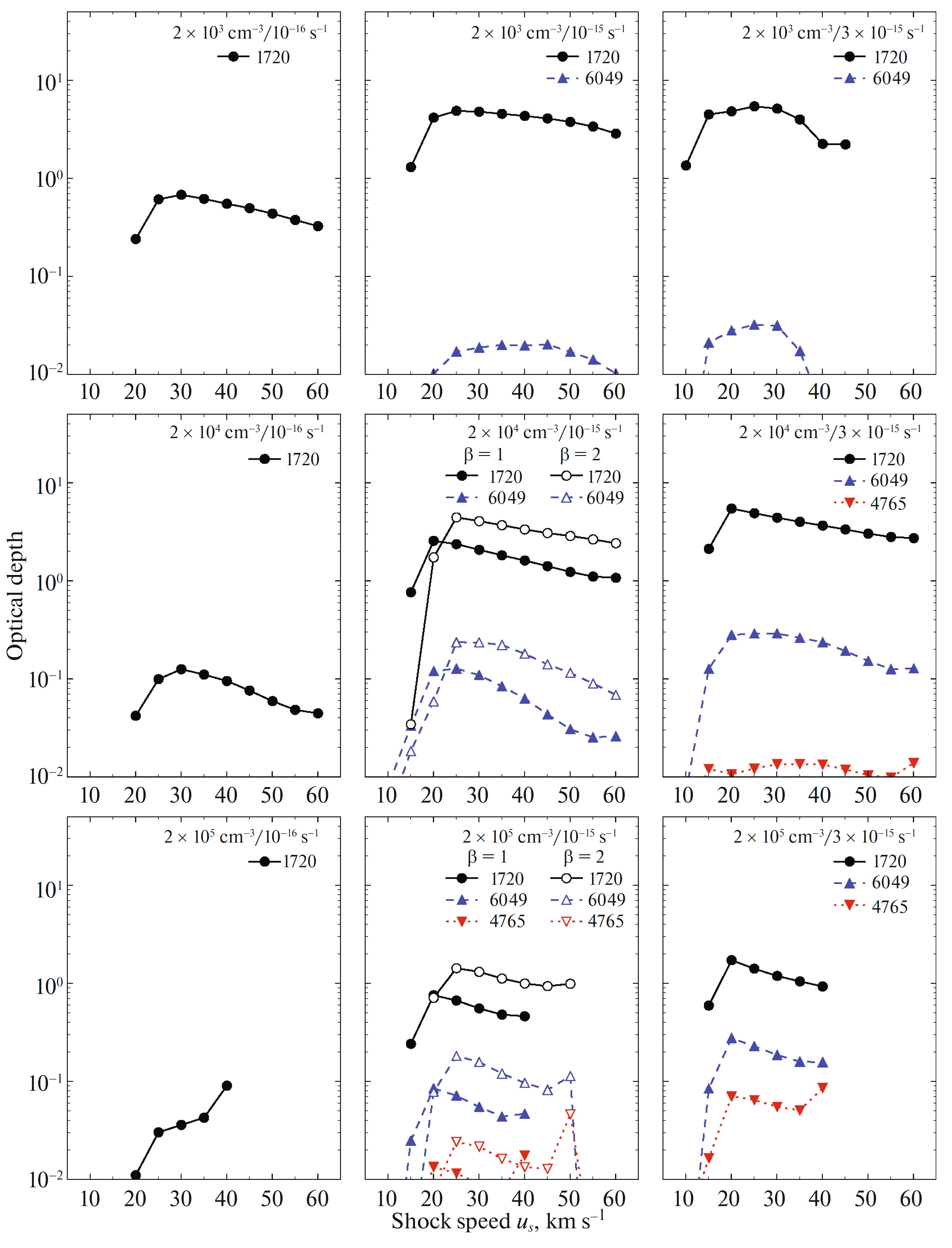}
\caption{Effective optical depth for OH maser transitions in the gas flow direction. The shock speed is along the horizontal axis. The values of the parameters $n_{\rm H,0}$ and $\zeta$ are indicated on each graph. The results of our calculations in which the preshock magnetic field strength was assumed to be twice as high, $\beta = 2$ (empty symbols), are also presented for $\zeta = 10^{-15}$~s$^{-1}$.}
\end{figure}

\section*{Discussion}
\subsection*{OH maser emission in SNRs}
The regions of 1720~MHz OH maser emission near SNRs have sizes of the order of 10$^{16}$~cm, with the most intense emission being generated in compact sources with sizes $\sim 10^{15}$~cm (Hoffman et al. 2005a, 2005b). The brightness temperature of the maser emission can reach $10^8 - 10^9$~K. The optical depth in the maser line along the line of sight must be about $15-20$ in this case (Hoffman et al. 2005b). For intense maser emission to be generated, the line of sight must be perpendicular to the gas flow direction (the shock is seen edge-on). The OH column density along the line of sight can exceed its value in the shock direction by several times, ${\rm cos}\theta \simeq 0.1$ (Lockett et al. 1999). Under this assumption the results of our calculations allow the observed intensities in the 1720 MHz line to be explained. At low initial densities, $n_{\rm H,0} = 2 \times 10^3$~cm$^{-3}$, the size of the postshock region, where the generation of intense OH maser emission is possible, is $\sim 10^{16}$~cm, which is greater than the observed sizes of the bright sources by an order of magnitude (Fig. 1d). Therefore, the bright OH maser sources are most likely clumps of relatively high density, $n_{\rm H_2} \simeq 10^5$~cm$^{-3}$.

So far no maser has been detected in the lines of excited OH rotational states in SNRs (Fish et al. 2007; Pihlstr\"{o}m et al. 2008; McDonnell et al. 2008). According to the estimates made by Pihlstr\"{o}m et al. (2008) and McDonnell et al. (2008), the optical depth along the line of sight in the 6049 and 4765~MHz lines needed for the detection of maser emission must exceed $2-3$. According to our calculations, the effective optical depth in these lines is $\tau_{\rm eff} \leq 0.3$ for all of the shock parameters under consideration. The optical depth along the line of sight is insufficiently high even under the assumption that ${\rm cos}\theta \simeq 0.1$. At high gas densities the 6049 and 4765~MHz masers are pumped more efficiently, but, at the same time, the shock length is smaller, which leads to low optical depths in this case as well.

The method of allowance for the spectral line overlap used in this paper is difficult to generalize to the simultaneous overlap of three lines -- in this case, the photon escape probability $\mathcal{P}$ in the expression for the intensity of radiation will depend on six parameters. The simultaneous overlap of three OH lines can affect the radiative transfer for a scatter of molecular velocities $v_{\rm D} > 0.3$~km~s$^{-1}$, which holds in our model. Therefore, the results obtained in this paper should be considered as an approximate estimate of the effect of line overlap on the collisional OH maser pumping mechanism. A higher-quality study of this effect is possible by the accelerated $\Lambda$-iteration method (Gray 2012).

\subsection*{OH maser emission in bipolar outflows near protostellar objects}

The process of star formation is accompanied by the ejection of gas flows by a young stellar object. Shocks are formed during the interaction of gas flows with the surrounding molecular cloud; the physical conditions in them could be favourable for the generation of OH maser emission (Litovchenko et al. 2012; de Witt et al. 2014). De Witt et al. (2014) undertook a search for the 1720~MHz OH maser emission toward 97 Herbig--Haro objects. No OH maser emission associated with bipolar outflows was detected. Bayandina et al. (2015) undertook a search for the OH maser emission toward 20 Extended Green Objects. These objects are most likely regions of massive star formation and high-speed outflows are located in them (Chen et al. 2013). The emission line at 1720 MHz was recorded only in one source and the emission region coincided with an ultracompact HII region within the positional uncertainty. Note that class I methanol masers are detected in bipolar outflows near protostellar objects and SNRs, which, just as the 1720 MHz maser, have the collisional pumping mechanism (see, e.g., Voronkov et al. 2006; McEwen et al. 2016).

A necessary condition for the generation of OH maser emission in the 1720 MHz line is a high gas ionization rate, $\zeta \gtrsim 10^{-15}$~s$^{-1}$. The gas ionization rate in molecular clouds near SNRs is estimated to be of the order of $10^{-15}$~s$^{-1}$ (Vaupr\'e et al. 2014; Shingledecker et al. 2016), while the gas ionization rates in molecular cloud cores and molecular outflows near protostellar objects are, as a rule, much lower (Caselli et al. 1998; Podio et al. 2014). An insufficiently high gas ionization rate can be responsible for the absence of OH maser emission in the 1720~MHz line in bipolar outflows near protostellar objects.

\section*{Conclusions}
We investigated the collisional pumping of OH masers in C-type shocks. The profiles of the gas temperature, velocity, and number densities of chemical species in the shock were computed using the numerical model published in Nesterenok (2018) and Nesterenok et al. (2019). In our calculations of the OH energy level populations we used the Sobolev approximation with continuum absorption and spectral line overlap. We showed that such parameters of the emission sources as the number density of OH molecules and the size of the region where the maser emission is amplified are not independent and are determined by the shock model. The largest optical depth in the 1720~MHz line is reached at high gas ionization rates $\zeta \geq 10^{-15}$~s$^{-1}$ and an initial density $n_{\rm H,0} \leq 2 \times 10^4$~cm$^{-3}$. There is also a population inversion in the lines of excited OH rotational states. However, the optical depth in these lines is small. The absence of OH maser emission in the 1720~MHz line in bipolar outflows near protostellar objects may be a consequence of a low gas ionization rate in these objects.

\section*{References}
 
\noindent
1. O.S. Bayandina, I.E. Val’tts, and S.E. Kurtz, Astron. Rep. {\bf 59}, 998 (2015).

\noindent
2. H. Beuther, A. Walsh, Y. Wang, M. Rugel, J. Soler, H. Linz, R.S. Klessen, L.D. Anderson, et al., Astron. Astrophys. {\bf 628}, A90 (2019).

\noindent
3. J. Le Bourlot, G. Pineau des For\^ets, D.R. Flower, and S. Cabrit, Mon. Not. R. Astron. Soc. {\bf 332}, 985 (2002).

\noindent
4. C.L. Brogan, D.A. Frail, W.M. Goss, and T.H. Troland, Astrophys. J. {\bf 537}, 875 (2000).

\noindent
5. C.L. Brogan, W.M. Goss, T.R. Hunter, A.M.S. Richards, C.J. Chandler, J.S. Lazendic, B.-C. Koo, I.M. Hoffman, et al., Astrophys. J. {\bf 771}, 91 (2013).

\noindent
6. V.V. Burdyuzha and D.A. Varshalovich, Sov. Astron. {\bf 17}, 308 (1973).

\noindent
7. P. Caselli, C.M. Walmsley, R. Terzieva, and E. Herbst, Astrophys. J. {\bf 499}, 234 (1998).

\noindent
8. J.L. Caswell, Mon. Not. R. Astron. Soc. {\bf 297}, 215 (1998).

\noindent
9. J.L. Caswell, Mon. Not. R. Astron. Soc. {\bf 308}, 683 (1999).

\noindent
10. X. Chen, C.-G. Gan, S.P. Ellingsen, J.-H. He, Z.-Q. Shen, and A. Titmarsh, Astrophys. J. Suppl. Ser. {\bf 206}, 9 (2013).

\noindent
11. R.M. Crutcher, Astrophys. J. {\bf 520}, 706 (1999).

\noindent
12. R.C. Doel, M.D. Gray, and D. Field, Mon. Not. R. Astron. Soc. {\bf 244}, 504 (1990).

\noindent
13. B.T. Draine, Astrophys. J. {\bf 241}, 1021 (1980).

\noindent
14. B.T. Draine and E.E. Salpeter, Astrophys. J. {\bf 231}, 77 (1979).

\noindent
15. M. Elitzur, Astrophys. J. {\bf 203}, 124 (1976).

\noindent
16. V.L. Fish, L.O. Sjouwerman, and Y.M. Pihlstr\"{o}m, Astrophys. J. {\bf 670}, L117 (2007).

\noindent
17. D.R. Flower and A. Gusdorf, Mon. Not. R. Astron. Soc. {\bf 395}, 234 (2009).

\noindent
18. D.A. Frail, W.M. Goss, and V.I. Slysh, Astrophys. J. {\bf 424}, L111 (1994).

\noindent
19. D.A. Frail and G.F. Mitchell, Astrophys. J. {\bf 508}, 690 (1998).

\noindent
20. I.E. Gordon, L.S. Rothman, C. Hill, R.V. Kochanov, Y. Tan, P.F. Bernath, M. Birk, V. Boudon, et al., J. Quant. Spectrosc. Radiat. Transfer {\bf 203}, 3 (2017).

\noindent
21. M.D. Gray, in Cosmic Masers -- from OH to H$_0$, Proceedings of the IAU Symposium {\bf 287}, Ed. by R.S. Booth, E.M.L. Humphreys, and W.H.T. Vlemmings (Cambridge Univ. Press, Cambridge, UK, 2012), p. 23.

\noindent
22. A. Gusdorf, S. Cabrit, D.R. Flower, and G. Pineau
des For\^ets, Astron. Astrophys. {\bf 482}, 809 (2008).

\noindent
23. A.N. Heays, A.D. Bosman, and E.F. van Dishoeck, Astron. Astrophys. {\bf 602}, A105 (2017).

\noindent
24. J.W. Hewitt, F. Yusef-Zadeh, M. Wardle, D.A. Roberts, and N.E. Kassim, Astrophys. J. {\bf 652}, 1288 (2006).

\noindent
25. J.W. Hewitt, F. Yusef-Zadeh, and M. Wardle, Astrophys. J. {\bf 706}, L270 (2009).

\noindent
26. I.M. Hoffman, W.M. Goss, C.L. Brogan, and M.J. Claussen, Astrophys. J. {\bf 620}, 257 (2005a).

\noindent
27. I.M. Hoffman, W.M. Goss, C.L. Brogan, and M.J. Claussen, Astrophys. J. {\bf 627}, 803 (2005b).

\noindent
28. D.G. Hummer and G.B. Rybicki, Astrophys. J. {\bf 293}, 258 (1985).

\noindent
29. F. Lique, P. Honvault, and A. Faure, Int. Rev. Phys. Chem. {\bf 33}, 125 (2014).

\noindent
30. I.D. Litovchenko, O.S. Bayandina, A.V. Alakoz, I.E. Val’tss, G.M. Larionov, D.V. Mukha, A.S. Nabatov, A.A. Konovalenko, V.V. Zakharenko, E.V. Alekseev, V.S. Nikolaenko, V.F. Kulishenko, and S.A. Odintsov, Astron. Rep. 56, {\bf 536} (2012).

\noindent
31. M.M. Litvak, Astrophys. J. {\bf 156}, 471 (1969).

\noindent
32. P. Lockett, E. Gauthier, and M. Elitzur, Astrophys. J. {\bf 511}, 235 (1999).

\noindent
33. S. Marinakis, Yu. Kalugina, and F. Lique, Eur. Phys. J. D {\bf 70}, 97 (2016).

\noindent
34. S. Marinakis, Yu. Kalugina, J. K{\l}os, and F. Lique,
Astron. Astrophys. {\bf 629}, A130 (2019).

\noindent
35. K.E. McDonnell, M. Wardle, and A.E. Vaughan, Mon. Not. R. Astron. Soc. {\bf 390}, 49 (2008).

\noindent
36. D. McElroy, C. Walsh, A.J. Markwick, M.A. Cordiner, K. Smith, and T.J. Millar, Astron. Astrophys. {\bf 550}, A36 (2013).

\noindent
37. B.C. McEwen, Y.M. Pihlstr\"{o}m, and L.O. Sjouwerman, Astrophys. J. {\bf 826}, 189 (2016).

\noindent
38. D.J. Mullan, Mon. Not. R. Astron. Soc. {\bf 153}, 145 (1971).

\noindent
39. A.V. Nesterenok, Mon. Not. R. Astron. Soc. {\bf 455}, 3978 (2016).

\noindent
40. A.V. Nesterenok, Astrophys. Space Sci. {\bf 363}, 151 (2018).

\noindent
41. A.V. Nesterenok, J. Phys.: Conf. Ser. {\bf 1400}, 022025 (2019).

\noindent
42. A.V. Nesterenok, J. Phys.: Conf. Ser. (2020, in
press), https://arxiv.org/abs/2009.06322.

\noindent
43. A.V. Nesterenok, D. Bossion, Y. Scribano, and F. Lique, Mon. Not. R. Astron. Soc. {\bf 489}, 4520 (2019).

\noindent
44. A.R. Offer, M.C. van Hemert, and E.F. van Dishoeck, J. Chem. Phys. {\bf 100}, 362 (1994).

\noindent
45. K.G. Pavlakis and N.D. Kylafis, Astrophys. J. {\bf 467}, 300 (1996).

\noindent
46. K.G. Pavlakis and N.D. Kylafis, Astrophys. J. {\bf 534}, 770 (2000).

\noindent
47. V.H.M. Phan, S. Gabici, G. Morlino, R. Terrier, J. Vink, J. Krause, and M. Menu, Astron. Astrophys. {\bf 635}, A40 (2020).

\noindent
48. Y.M. Pihlstr\"{o}m, V.L. Fish, L.O. Sjouwerman, L.K. Zschaechner, P.B. Lockett, and M. Elitzur, Astrophys. J. {\bf 676}, 371 (2008).

\noindent
49. L. Podio, B. Lefloch, C. Ceccarelli, C. Codella, and R. Bachiller, Astron. Astrophys. {\bf 565}, A64 (2014).

\noindent
50. S.S. Prasad and S.P. Tarafdar, Astrophys. J. {\bf 267}, 603 (1983).

\noindent
51. W.H. Press, S.A. Teukolsky, W.T. Vetterling, and B.P. Flannery, Numerical Recipes: The Art of Scientific Computing (Cambridge Univ. Press, New York, 2007).

\noindent
52. H.-H. Qiao, S.L. Breen, J.F. G\'omez, J.R. Dawson, A.J. Walsh, J.A. Green, S.P. Ellingsen, H. Imai, et al., Astrophys. J. Suppl. Ser. {\bf 247}, 5 (2020).

\noindent
53. W.G. Roberge and B.T. Draine, Astrophys. J. {\bf 350}, 700 (1990).

\noindent
54. M. Ruaud, V. Wakelam, and F. Hersant, Mon. Not. R. Astron. Soc. {\bf 459}, 3756 (2016).

\noindent
55. F.L. Sch\"{o}ier, F.F.S. van der Tak, E.F. van Dishoeck, and J.H. Black, Astron. Astrophys. {\bf 432}, 369 (2005).

\noindent
56. F. Schuppan, C. R\"{o}ken, and J. Becker Tjus, Astron. Astrophys. {\bf 567}, A50 (2014).

\noindent
57. C.N. Shingledecker, J.B. Bergner, R. Le Gal, K.I. \"{O}berg, U. Hincelin, and E. Herbst, Astrophys. J. {\bf 830}, 151 (2016).

\noindent
58. V.V. Sobolev, Sov. Astron. {\bf 1}, 678 (1957).

\noindent
59. S. Vaupr\'e, P. Hily-Blant, C. Ceccarelli, G. Dubus,
S. Gabici, and T. Montmerle, Astron. Astrophys. {\bf 568}, A50 (2014).

\noindent
60. M.A. Voronkov, K.J. Brooks, A.M. Sobolev, S.P. Ellingsen, A.B. Ostrovskii, and J.L. Caswell, Mon. Not. R. Astron. Soc. {\bf 373}, 411 (2006).

\noindent
61. M. Wardle, Astrophys. J. {\bf 525}, L101 (1999).

\noindent
62. M. Wardle and K. McDonnell, in Cosmic Masers -- from OH to H$_0$, Proceedings of the IAU Symposium {\bf 287}, Ed. by R.S. Booth, E.M.L. Humphreys, and W.H.T. Vlemmings (Cambridge Univ. Press, Cambridge, UK, 2012), p. 441.

\noindent
63. M. Wardle and F. Yusef-Zadeh, Science (Washington, DC, U.S.) {\bf 296}, 2350 (2002).

\noindent
64. A. de Witt, M. Bietenholz, R. Booth, and M. Gaylard, Mon. Not. R. Astron. Soc. {\bf 438}, 2167 (2014).

\noindent
65. F. Yusef-Zadeh, M. Wardle, J. Rho, and M. Sakano, Astrophys. J. {\bf 585}, 319 (2003).

{\it Translated by V. Astakhov}

\end{document}